\documentclass[11pt]{article}
\usepackage{latexsym}
\usepackage{amsmath}
\usepackage{epsfig}
\usepackage{color}
\usepackage{graphicx}
\usepackage{amsfonts}
\usepackage{amssymb}
\usepackage{rotating}
\newcommand{\be}{\begin{equation}}
\newcommand{\ee}{\end{equation}}
\newcommand{\ben}{\begin{eqnarray}}
\newcommand{\een}{\end{eqnarray}}

\numberwithin{equation}{section}

\begin{document}

\title{{Bayesian outlier detection in Capital Asset Pricing Model}}
\author{Maria Elena De Giuli \and Mario Alessandro Maggi \and
Claudia Tarantola \thanks{Address for correspondence: Claudia
Tarantola, Department of Economics and Quantitative Me\-thods ,
Via S. Felice 7, 27100
Pavia, Italy. \texttt{E-mail:~claudia.tarantola@unipv.it}}\\
Department of Economics and Quantitative Methods, University of
Pavia, Italy }

\date{}
\maketitle
\begin{abstract}

 We propose a novel Bayesian optimisation procedure for
outlier detection in the Capital Asset Pricing Model.  We use a
parametric product partition model to robustly estimate the
systematic risk of an asset. We assume that the returns follow
independent normal distributions and we impose a partition
structure on the parameters of interest. The partition structure
imposed on the parameters induces a corresponding clustering of
the returns. We identify
 via an
optimisation procedure the partition that best separates standard
observations from the atypical ones. The methodology is
illustrated with reference to a real data set, for which we also
provide a microeconomic interpretation of the detected outliers.

\end{abstract}


\section{Introduction}

In this paper we propose a novel Bayesian optimisation procedure
for outlier identification in a Capital Asset Pricing framework.
The Capital Asset Pricing Model (CAPM), see Sharpe~(1964),
Lintner~(1965), Mossin~(1966) and Black~(1972), states that an
asset expected return is equal to the risk free rate plus a prize
for risk. The CAPM is  widely used in applications to evaluate the
performance of assets and portfolios, and different performance
measures are based on it, see e.g. chap. 4 in Amenec and Le Sourd
(2003). In particular, it is very useful for the calculation of
the cost of capital equities, which is necessary for market based
firm value models, see e.g. Ross et al.~(2008). The CAPM can be
represented by a simple linear regression where the slope
identifies the systematic risk of an asset, that is  it measures
the  return sensitivity to movement in the market. The systematic
risk represents the component of the risk that cannot be
eliminated simply via portfolio diversification. Therefore,
systematic risk is a key variable to be taken into account for
asset allocation and portfolio management.

Almost all empirical analysis of the CAPM has been carried out in
the classical framework, see i.e.\ Fama and French~(2004) for an
exaustive review. The systematic risk is usually estimated by the
least square method which coincides with the maximum likelihood
estimator under the assumption of normality. It is well known that
this approach has at least two disadvantages.  Firstly, this
estimation method is sensitive to the presence of  outliers, that
is observations that do not follow the same statistical model  as
the main part of the data. Secondly, it is not possible  to
incorporate prior beliefs about behaviour of  returns  in the
model.

In this paper, to overcome these problems,  we focus on Bayesian
robust estimation procedures for  linear regression models; see
e.g. Chaturvedi~(1996), Fern\'{a}ndez et al.~(2001), Quintana and
Iglesias~(2003) and Quintana et al.~(2005a, 2005b). In particular,
Quintana and Iglesias~(2003) and Quintana et al.~(2005a) show that
outlying points can be accommodated either by a product partition
 model with a normal structure on the returns,  or by a simple regression model with $t$ shape
errors and small  degrees of freedom.

We follow the first approach and we apply  a normal model with a
partition structure on the parameters of interest. This approach
 has at least two advantages. Firstly, the use of a normal
distribution is consistent with the assumption of  mean-variance
analysis required by CAPM. Secondly, the use of a partition
structure simultaneously yields outlier identification and model
robustification. The partition structure allows us to separate the
main body of ``standard'' data points from the ``atypical'' ones.

 Regarding the outlier identification problem we work in  a Bayesian decision the\-o\-re\-ti\-cal
 framework.
 The partition  that best separates ``standard'' observations from the
 ``atypical'' ones is selected by
  minimising a specific score function. In Quintana and
Iglesias~(2003) this partition is identified by applying a
clustering algorithm. However, as they mentioned  (see page 572 in
Quintana and Iglesias,~2003), a weakness of their algorithm is
that it could be trapped in local modes. Furthermore, as they
select outlying points one by one, they could incur in the problem
of {masking}. If  a data set has multiple outliers, they may mask
one another, making outlier identification difficult. Masked
outliers should be removed as a group, otherwise
 their presence could remain undetected.
 We overcome these problems  by applying  a
constrained optimisation algorithm to select the optimal
partition. Our algorithm includes a
 preliminary step in which the data are prescreened
via a robust technique that allows us to identify a set of
potential outliers. Subsequently, outliers are efficiently
selected among the potential ones.

The paper is structured as follows. In Section~2 we briefly
introduce the CAPM and  parametric product partition models (PPM).
In Section~3 we describe the optimisation algorithm used for
outlier identification. For comparative purposes, in Section~4 we
apply our procedure  to the IPSA data  set,
 previously examined by Quintana et al.~(2005a).  Conclusions are given in  Section~5.

\section{Background and preliminaries \label{back}}

{\subsection{The CAPM}}

  According to CAPM,  the  expected return of
 any
asset $i$  is a linear function of the market portfolio one
\begin{equation}
E\left(R_{i}\right)
=R_{f}+\beta_{i}\left(E\left(R_{m}\right)-R_{f}\right), \quad
i=1,\ldots,N\label{capm0},
\end{equation}

\noindent where $N$ is the number of assets,  $R_{i}$ is the
return of asset $i$, $R_{f}$ is the risk-free rate of return,
$R_{m}$ is the return on the market portfolio  and the slope
$\beta_i$ measures the systematic risk. The market portfolio is
the portfolio containing every asset available to the economic
agent, in amounts proportional to their total market values.
Market portfolio is  a theoretical concept, hence it is necessary
to use a market index return, $R_{M}$, as an observable proxy of
the market portfolio return $R_{m}$.

An  extension of equation  (\ref{capm0}) is often used to estimate
the systematic risk.  Another coefficient denoted
 by $\alpha_i$ is usually introduced, obtaining the
following expression
 \begin{equation}
E\left(R_{i}\right)-R_{f}
=\alpha_{i}+\beta_{i}\left(E\left(R_{M}\right)-R_{f}\right), \quad
i=1,\ldots,N
 \label{capm2},
\end{equation}

\noindent where   $\alpha_i$ denotes wether the asset $i$
over/under-performs the expected return explained by CAPM. The
parameters of equation (\ref{capm2}) are estimated by using the
linear regression equation \vspace{-0.6cm}
\begin{eqnarray}
&\;&R_{it}-R_{ft}
=\alpha_{i}+\beta_{i}\left(R_{Mt}-R_{ft}\right)+\varepsilon_{it},\nonumber\\
&\;&y_{it}    =\alpha_{i}+\beta_{i}x_{t}+\varepsilon_{it}, \quad
i=1,\ldots,N,\quad t=1,\ldots,T\label{capm3},
\end{eqnarray}
where, for a  $t$-period and a generic asset $i$, $y_{it}$ is the
excess return of the asset, $R_{it}$ denotes the return of the
asset, $R_{ft}$ is the risk-free rate of return,
 $R_{Mt}$ is the market
index return and $\varepsilon_{it}$ is a normally distributed
error term.

 The estimation of the systematic risk $\beta_i$ can be
affected by the presence of outlying points.  An approach that
carefully takes into account
  the presence
of anomalous observations should be applied.  In this paper we
model outliers  by a shift in the regression mean and we handle
them working with an extension of equation (\ref{capm3}). More
precisely, we assume that excess returns of share $i$, at
different time points $t$, can be more appropriately described by
a set of parallel regression lines (see also Quintana and
Iglesias,~2003 and Quintana et al.,~2005a). We allow $\alpha_i$ to
change with $t$ (indicated as $\alpha_{it}$), and we estimate the
following equation
\begin{equation}y_{it} =\alpha_{it}+\beta_{i}x_{t}+\varepsilon_{it},
\quad i=1,\ldots,N,\quad t=1,\ldots,T, \label{capm4}
\end{equation}
where $\alpha_{it}$ assumes values in the finite  set
\mbox{$\boldsymbol{\alpha}_{\rho_i}=\left(\alpha^{*}_{i1},\ldots,\alpha^{*}_{i|\rho_i|}\right)$},
with cardinality $|\rho_i|$ smaller then $T$  (more details will
be provided in Section  \ref{PPMs}).
 In fact,  it can be  realistically assumed that the number of  regression
 lines is inferior to the number of observations. Our aim is to group together (to cluster)
time periods with common values of the intercepts. These groups
 lead to
 a clustering of the corresponding excess returns $y_{it}$. We will end up with a main group of standard observations
 and one or more groups of atypical ones.
The number and the composition of the groups (\emph{the partition
structure}) is unknown,  hence we assign a prior distribution to
the set of all possible partitions, see Section \ref{PPMs} for the
details.

\subsection{A parametric product partition model  \label{PPMs}}

 Following Quintana and Iglesias~(2003)
and Quintana et al.~(2005a), we use a parametric product partition
model (PPM)
 to robustly estimate the systematic
risk and to identify outlying points. We now briefly review the
theory on parametric product partition models  with reference to
our specific problem, see Barry and Hartigan~(1992) for a detailed
and more general presentation.

Given the model described by equation (\ref{capm4}), let
$S_i^{0}=\{1, \ldots,T \}$  be the set of  time periods. A
partition of the  set $S_i^0$,  $\rho_i = \left\{ S_i^{1}, \ldots,
S_i^{d}, \ldots, S_i^{|\rho_i|} \right\}$ with cardinality
$|\rho_i|$,  is defined by the property that $S_i^d \cap
S_i^{d^{\prime}}=\emptyset$ for $d \neq d^{\prime}$ and $\cup_d\,
S_i^d=S_i^0$. The generic element of $\rho_i$ is $S_i^{d}=\left\{
t: \alpha_{it}=\alpha^{*}_{id}\right\}$, where
\mbox{$\boldsymbol{\alpha}_{\rho_i}=\left(\alpha^{*}_{i1},\ldots,\alpha^{*}_{i|\rho_i|}\right)$}
is the vector of the unique  values of
$\boldsymbol{\alpha}_i=\left(\alpha_{i1},\ldots,\alpha_{iT}\right)$.
All $\alpha_{it}$ whose
 subscripts $t$ belong to the same set
 $S_i^d \in \rho_i$ are (\emph{stochastically})
equal; in this sense they are regarded as a single \emph{cluster}.

We assign to each partition $\rho_i$ the following  prior
probability

\begin{equation}
P\left(\rho_i=\left\{ S_i^{1}, \ldots,S_i^{|\rho_i|}
\right\}\right)=K \prod_{d=1}^{|\rho_i|} C\left(S_i^{d}\right),
\label{product distribution rho}
\end{equation}
where $C\left(S_i^{d}\right)$ is a \emph{cohesion function} and
$K$ is the normalising constant.  Equation (\ref{product
distribution rho}) is referred to as the \emph{product
distribution} for partitions. The cohesions represent prior
weights on group formation and  formalise our opinion on how
tightly clustered the elements of $S_i^{d}$ would be.

The cohesions can be specified in different ways, a useful choice
is \ben \label{eqcohesion1} C\left(S_i^d\right)= c \times
\left(\left|S_i^d\right|-1\right)!, \quad \quad
d=1,\ldots,|\rho_i|\een for some positive constant $c$.

For moderate values of $c$, e.g.\ $c=1$, the cohesions in
(\ref{eqcohesion1}) yield a prior distribution that favours the
formation of partitions with a reduced number of large subsets.
This is a desirable feature for an outlier detection model, since
we do not want to identify too many subsets of points as outliers.
 For more
details on the choice of $c$ see i.e. \ Liu~(1996), Quintana et
al.~(2005b) and Tarantola et al.~(2008).

Moreover, there is an interesting connection between parametric
PPMs and the class of Bayesian nonpara\-me\-tric models based on a
mixture of Dirichlet Processes (Antoniak,~1974).  Under the latter
prior, the marginal distribution of the observables is a specific
PPM with the cohesion functions specified by
equation~(\ref{eqcohesion1}), see Quintana and Iglesias~(2003).
This connection allows us  to use efficient Markov Chain Monte
Carlo (MCMC) algorithms developed for Bayesian nonparametric
problems like the one that we apply here.

In this paper we consider  the following Bayesian hierarchical
model
\begin{align*}
&  y_{it}|\rho_i,
\left(\alpha^{*}_{i1},\ldots,\alpha^{*}_{i|\rho_i|}\right),\beta_i,\sigma_i^{2}
\overset{ind}{\sim}N\left(  \alpha_{it}+\beta_i x_{t}, \sigma_i^{2}\right) \\
&
\alpha^{*}_{i1},\ldots,\alpha^{*}_{i|\rho_i|}|\rho_i,\sigma_i^{2}
\overset
{IID}{\sim}N \left(  a,\tau_{0}^{2} \sigma_i^{2} \right) \\
&  \beta_i|\sigma_i^{2} \sim N \left(  b,\gamma_{0}^{2} \sigma_i^{2}\right) \\
&  \rho_i\sim\mbox{product distribution,}\;\mbox{with}\;
C\left(S_i^d\right)= c \times
\left(\left|S_i^d\right|-1\right)!\\
&  \sigma_i^{2} \sim IG(v_{0},\lambda_{0}),
\end{align*}
where
 $a$, $b$, $\tau^{2}_{0}$, $\gamma^{2}_{0}$,
$v_{0}$ and $\lambda_{0}$ are user-specified hyperparameters, the
product distribution is defined in~(\ref{product distribution
rho}) and $IG\left(v_{0},\lambda_{0}\right)$ is an inverted gamma
distribution with
$E\left(\sigma_i^2\right)=\lambda_0/\left(v_0-1\right)$.  The
Gibbs algorithm applied to sample from the posterior distributions
of the parameters  is described in the Appendix.

\section{Optimal outlier detection \label{optimisation}}

 To  detect outlying
points we apply   a constrained optimisation algorithm, working in
a Bayesian decision theoretic framework. Our aim is to select the
partition that best separates the main group of standard
observations from one or more groups of aty\-pi\-cal data. Each
partition corresponds to a different model and the best model is
the one minimising a given loss function. We consider a loss
function that combines the estimation of the parameters  and the
partition selection problems.

Given a generic asset $i$, let
$\left(\boldsymbol{\alpha}_i,\beta_i,\sigma^2_i\right)$  be the
vector of  parameters of the model  and
$\left(\boldsymbol{\alpha}_{\rho_i},\beta_{\rho_i},\sigma^2_{\rho_i}\right)$
be the corresponding vector  that results when fixing $\rho_i$. We
consider the loss function
\begin{eqnarray}
L\left(\rho_i,\boldsymbol{\alpha}_{\rho_i},\beta_{\rho_i},\sigma^2_{\rho_i},
\boldsymbol{\alpha}_i,\beta_i,\sigma^2_i\right)=\frac{k_1}{T}\
\parallel\boldsymbol{\alpha}_{\rho_i}-\boldsymbol{\alpha}_i\parallel^2\ +\ k_2\
\left(\beta_{\rho_i}-\beta_i\right)^2\ +\nonumber\\
+ k_3\ \left(\sigma^2_{\rho_i}-\sigma_i^2\right)^2\ +\
\left(1-k_1-k_2-k_3\right)|\rho_i|,\label{loss}
\end{eqnarray}
where  $\parallel \cdot \parallel$ is the Euclidean norm, and
$k_j$ ($j=1,2,3$) are positive cost-complexity parameters with
$\sum_{j=1}^3 k_j\leq 1$. Minimizing the expected value
of~(\ref{loss}) is equivalent to choosing the partition %$\rho_i^*$
that minimises the following score function
\begin{eqnarray}
SC\left(\rho_i\right)=\frac{k_1}{T}\
\parallel\hat{\boldsymbol{\alpha}}_{i}^B(y)-\hat{\boldsymbol{\alpha}}_{\rho_i}(y)\parallel^2\ +\ k_2\
\left[\hat{\beta}_{i}^B(y)-\hat{\beta}_{\rho_i}(y)\right]^2\
+\nonumber\\+ k_3\
\left[\big(\hat{\sigma}^{B}_{i}(y)\big)^2-\hat{\sigma}_{\rho_i}^2(y)\right]^2\
+\ \left(1-k_1-k_2-k_3\right)|\rho_i|.\label{score}
\end{eqnarray}
 In~(\ref{score}), a superscript ``$B$'' means that we consider the
Bayesian estimates of the corresponding parameter, whereas a
subscript ``$\rho_i$'' denotes the estimate of the parameter (or
vector of parameters) conditionally on the partition $\rho_i$.
Formally, if we indicate with  $\theta$  a generic parameter
in~(\ref{score}), we get $\hat{\theta}^B(y)=E(\theta|y)$ and
$\hat{\theta}_{\rho_i}(y)=E(\theta|y,\rho_i)$.
 The  estimates
$\hat{\theta}^B(y)$ and $\hat{\theta}_{\rho_i}(y)$ of $\theta$ are
obtained via the MCMC method described in the Appendix.

The number of all possible partitions is equal to $B(a)$, the
\emph{Bell number} of order $a$,  recursively defined by
$B(a+1)=\sum_{k=0}^a \binom{a}{k}B(k)$, with $B(0)=1$. This
quantity is extremely large even for moderate values of $a$,
therefore we  need to restrict our search to a tractable subset of
all partitions.

To avoid evaluating and comparing the scores of an impossibly
large number of partitions,   we propose a two step algorithm.
This algorithm reduces by construction the probability of
incurring the masking problem.
 It examines all partitions having a given
structure, and groups of observations may be included/excluded as
block in the different clusters.

 In the
first step of the algorithm we use least trimmed squares (LTS)
regression, see Rousseeuw~(1984) and Rousseeuw and Leroy~(1987),
to prescreen the data and identify a large set of potential
outliers. A similar idea has been successfully applied by Hoeting
et al.~(1996) for simultaneous variable selection and outlier
identification in a linear regression model.  LTS  is also used in
the Bayesian Model Averaging Package of R (Raftery et al.,~ 2008)
to prescreen the data.

   In the
second step, we constrain our search to partitions identifying as
outliers only particular subsets of those identified by LTS, and
we select the one that minimises the score function
($\ref{score}$).

Among robust techniques, we have chosen  LTS since it has a very
high ({finite-sample}) breakdown point (close to $1/2$) and tends
to identify a large number of observations as abnormal, reducing
the possibility of misclassifying anomalous points. However, it
should be noticed that LTS can be rather sensitive to small
perturbations in the central part of the data (high subsample
sensitivity), see e.g. Ellis (1998), V\'{i}\v{s}ek (1999), and
\v{C}\'{i}\v{z}ek and  V\'{i}\v{s}ek (2000). Attention should be
paid to check if the set of  potential outliers is reasonable. In
the specific case examined here the set of potential outliers is
sensible. In fact the elements selected by LTS correspond to
``small/high'' values of the components
  $\widehat{\boldsymbol{\alpha}}^B_{i}(y)$.

\color{black}

The algorithm consists of the following two steps. Let $i$ be a
generic asset.

 \begin{itemize}

 \item[\emph{Step 1.}]
We apply  LTS to  the excess returns of the asset $i$. All points
with an absolute value of the standardized residuals greater than
$2.5$
 are considered as potential outliers.

 \item[\emph{Step 2}.]

 Let  $\mathcal{O}^{LTS}_i$ be the
set of all time points corresponding to the potential outliers
identified in Step 1.
 We
restrict our search to partitions with  cardinality $2$   or $3$
 where the outliers are
particular subsets of $\mathcal{O}^{LTS}_i$.

The  data are classified in  clusters $S_i^1$, $S_i^2$ and $S_i^3$
defined as follows. Cluster $S_i^2$  contains  ``standard
observations'', with $\left(S^0_i\setminus
O_i^{LTS}\right)\subseteq S_i^2$. The remaining data, identifying
``anomalous points'',  are classified either in $S_i^1$
(``anomalous low values") or in $S_i^3$ (``anomalous high values")
with $\left(S_i^1\cup S_i^3\right)\subseteq O_i^{LTS}$.

 We only consider
partitions  with the following alternative structures
$\rho_i=\left\{S_i^1,S_i^2,S_i^3\right\}$ or $\rho_i=\left\{S_i^2,
\left(S_i^1 \cup S_i^3\right)\right\}$. If a cluster is empty it
is not considered as a component of $\rho_i$. We do not consider
the case in which both $S_i^1$ and $S_i^3$ are empty.

The  methodology used to construct  the clusters $S_i^1$, $S_i^2$
and $S_i^3$ is described below.

\begin{itemize}

\item[i)] Given the vector
$\widehat{\boldsymbol{\alpha}}^B_{i}(y)=\left(\widehat{{\alpha}}^B_{i_1},\ldots,
\widehat{{\alpha}}^B_{i_T}\right)$ of the Bayesian estimates of
the intercepts of model (\ref{capm4}), we indicate with $me$ the
median of its elements.

\item[ii)] For each time point $\ell \in \mathcal{O}^{LTS}_i$ we
compute the  deviation from the median,
$d_\ell=\left(\widehat{{\alpha}}^B_{i_\ell}-me\right)$.

\item[iii)] We construct the set $\mathcal{D}$ containing all
deviations from the median and two instrumental extra points
$$\mathcal{D}=\left\{\left\{d_\ell, \;\; \ell \in
\mathcal{O}^{LTS}_i\right\} \bigcup \left\{ \min\limits_\ell
\left(d_\ell\right)-\kappa,\  \max\limits_\ell
\left(d_\ell\right)+\kappa\right\}\right\}, $$

with $\kappa>0$ such that $\min\limits_\ell \left(d_\ell\right)-\kappa<0$ and
${\max\limits_\ell \left(d_\ell\right)}+\kappa> 0. $

 \item[iiii)] For every possible couple of values
$\left(d^L,d^U\right) \in \left(\mathcal{D}\times
\mathcal{D}\right)$, with \mbox{$ d^L< 0$} and $0\leq d^U$, we
classify in $S_i^1$ all $\ell \in \mathcal{O}^{LTS}_i$ such that
$d_\ell\leq d^L$ and in $S_i^3$ all $\ell \in \mathcal{O}^{LTS}_i$
that $d_\ell\geq d^U$, that is $S_i^1=\left\{\ell \in
\mathcal{O}^{LTS}_i: d_\ell \leq d^L\right\}$ and
$S_i^3=\left\{\ell \in \mathcal{O}^{LTS}_i: d_\ell\geq
d^U\right\}$. The remaining points are classified in $S_i^2$.

In this way we construct a list of possible partitions, that will
then be compared in terms of the  value of the score function
(\ref{score}). The optimal partition is the one with  the minimum
 score function value.

\end{itemize}

\end{itemize}

\section{ Analysis of the IPSA stock market data \label{chil}}

To test the performance of our  procedure we analysed the IPSA
stock market data, previously examined  by Quintana et
al.~(2005a). The IPSA  is the main index of the ``Bolsa de
Comercio de Santiago'' (Santiago Stock Exchange). It corresponds
to a portfolio containing  the 40 most heavily traded stocks, the
list is revised quarterly.

We considered monthly data  relative to the period January
1990-June 2004. We used the IPSA index as a proxy of the Chilean
market portfolio and the interest rate of Central Bank discount
bonds as the risk free rate. We focused our analysis only on the 5
shares listed in Table~\ref{Tab1}, for which Quintana et
al.~(2005a) provided  a detailed analysis  both of the estimates
of the parameters  and of the selected partitions.

 We used the following values of
the hyperparameters  $c=1$, $a=0$, $b=1$,
$\tau_0^2=\gamma_0^2=1000$, $v_0=2.0001$, $\lambda_0=0.010001$. We
set  $\left(k_1, k_2, k_3\right)=\frac{1}{2012}(1000,1000,1)$
in~(\ref{loss}) and~(\ref{score}), to give priority to the
estimation of $\boldsymbol{\alpha}_i$ and $\beta_i$, imposing weak
restriction on the number of clusters. These values lead to the
same prior distributions and the same relative weights for the
score function components as  in Quintana et al.~(2005a).

The two MCMC algorithms, used respectively to obtain the Bayesian
estimates of the parameters and the estimates given a specific
partition, are both based on  a run of $10\thinspace 000$ sweeps
with a burn-in of $1\thinspace 000$ iterations. Convergence of the
MCMC was assessed using  standard  criteria, see e.g.\ Best et
al.~(1995) and Cowles and Carlin~(1996). No specific indication of
abnormal behaviour is obtained. The two MCMC algorithms required
16 and 2.9 minutes respectively per $10\thinspace 000$ iterations
on a Pentium IV 3.4 GHz, 1 GB RAM personal computer.
 The programs were written in MATLAB; it is expected that
a lower level programming language could speed up the execution
time by a factor of at least $5$.

\subsection{Numerical results}

In Table~\ref{Tab1} we report the partitions selected by our
algorithm   and the one proposed by Quintana et al.~(2005a),
denoted by   DMT and QIB respectively. The MCMC standard errors of
the estimates were calculated by splitting the Markov chain output
into batches, see Geyer~(1992). It is also indicated  the value of
the  score function $SC$ and of the proportional reduction in
score ($PRS$),  that is
$PRS=(\mathrm{SC}_\mathrm{QIB}-\mathrm{SC}_\mathrm{DMT})/\mathrm{SC}_\mathrm{QIB}$;
for all  shares $SC_{DMT}\leq SC_{QIB}$. The PRS is  large for $3$
out of $5$ cases (more than $33\%$), in particular it is equal to
 $61.97\%$ for the Concha y Toro share.

\begin{center}
TABLE \ref{Tab1} ABOUT HERE \end{center}

 Note
that for the Concha y Toro  and Entel shares  the outliers
selected by Quintana et al.~(2005a) are a subset of those
identified with our procedure. For the  Cementos B\'io B\'io S.A.\
and Copec S.A.\ we select the same outlier set, but since we group
them in only one cluster we obtain a lower value of the score
function.

\begin{figure}[p]\caption{Score values for the different partition explored by
the algorithm of DMT. The optimal value is indicated by a star. A
cross represents the  partition selected by QIB.\medskip}
\includegraphics[height=5.8cm]{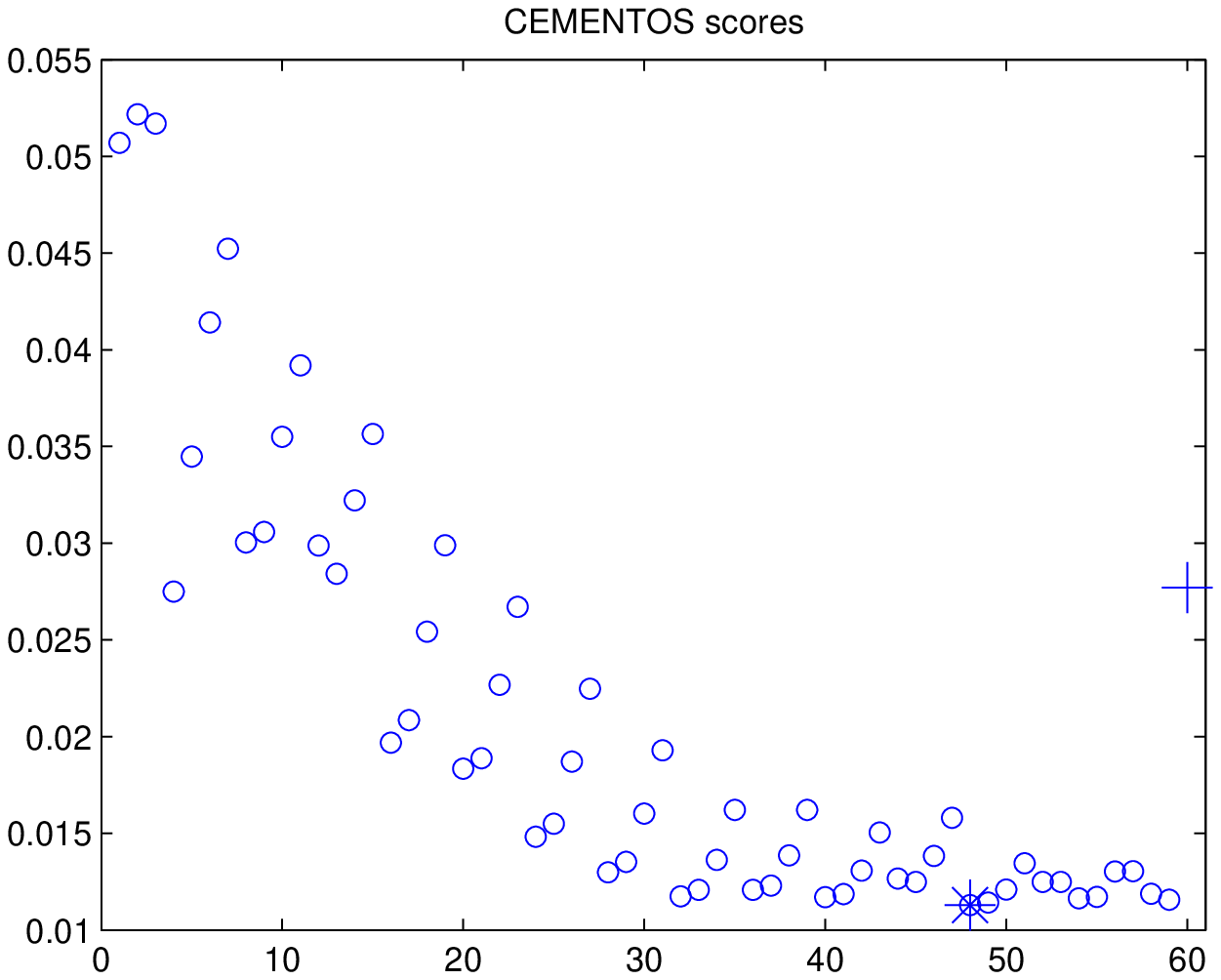}
\includegraphics[height=5.8cm]{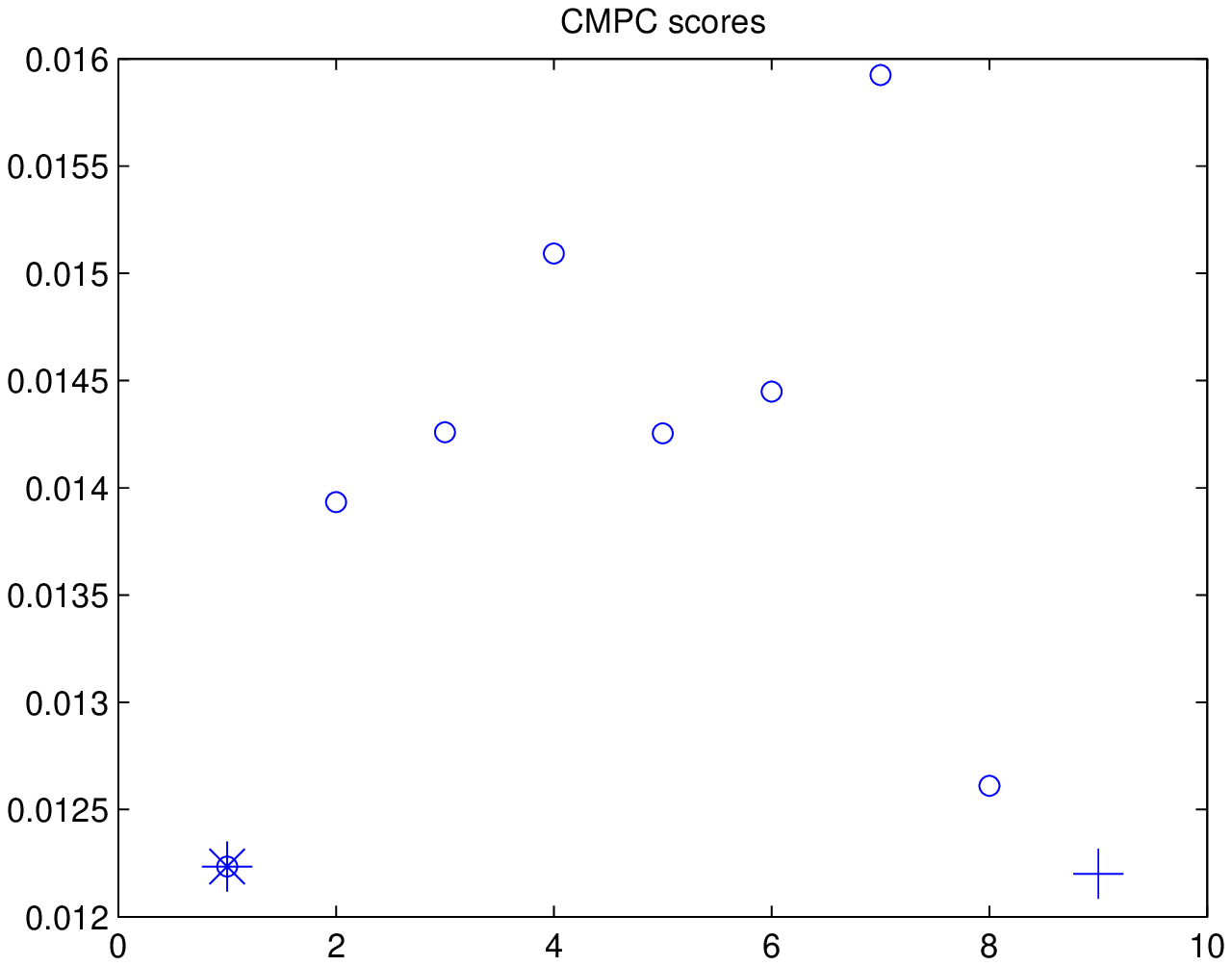}\\
\includegraphics[height=5.8cm]{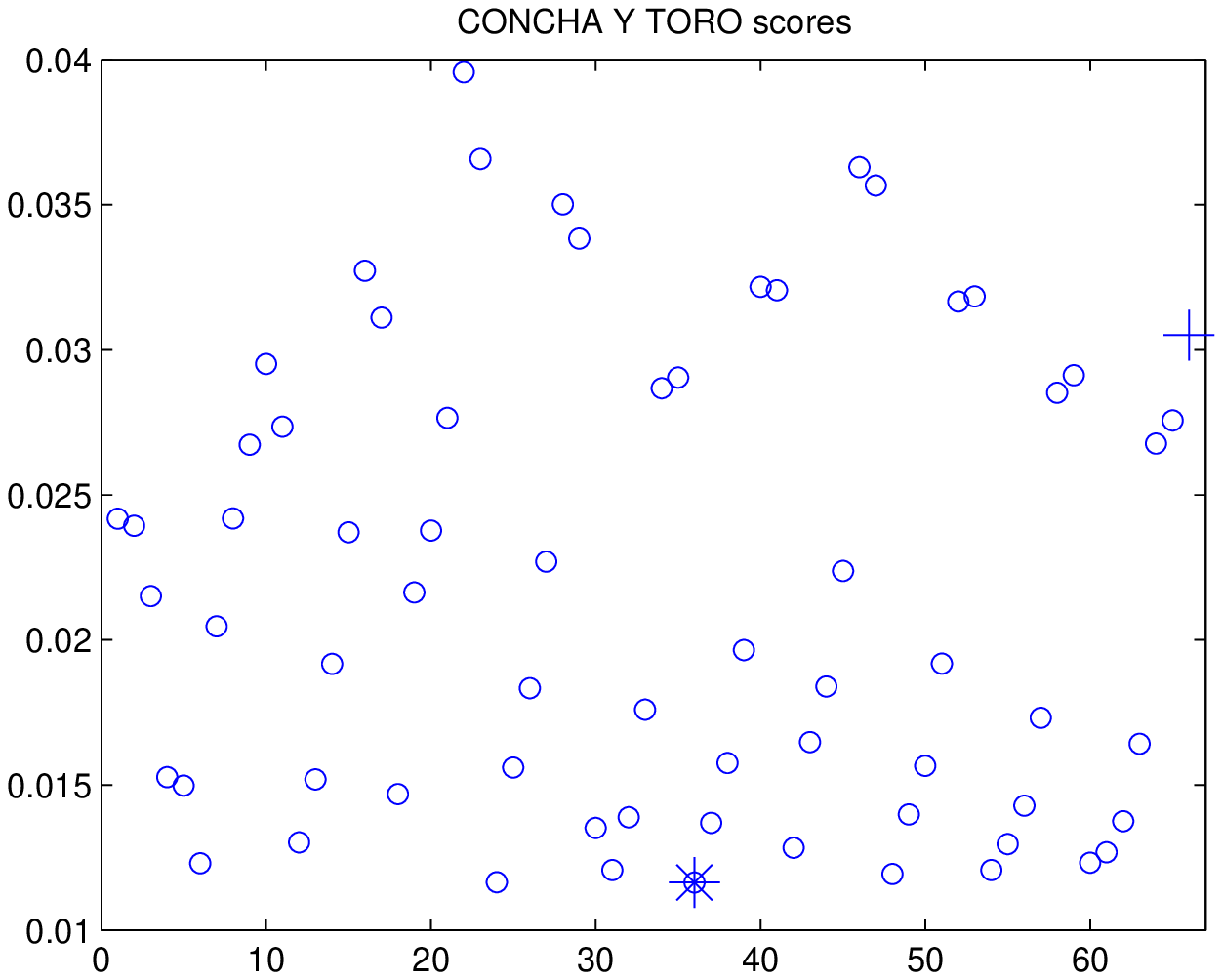}
\includegraphics[height=5.8cm]{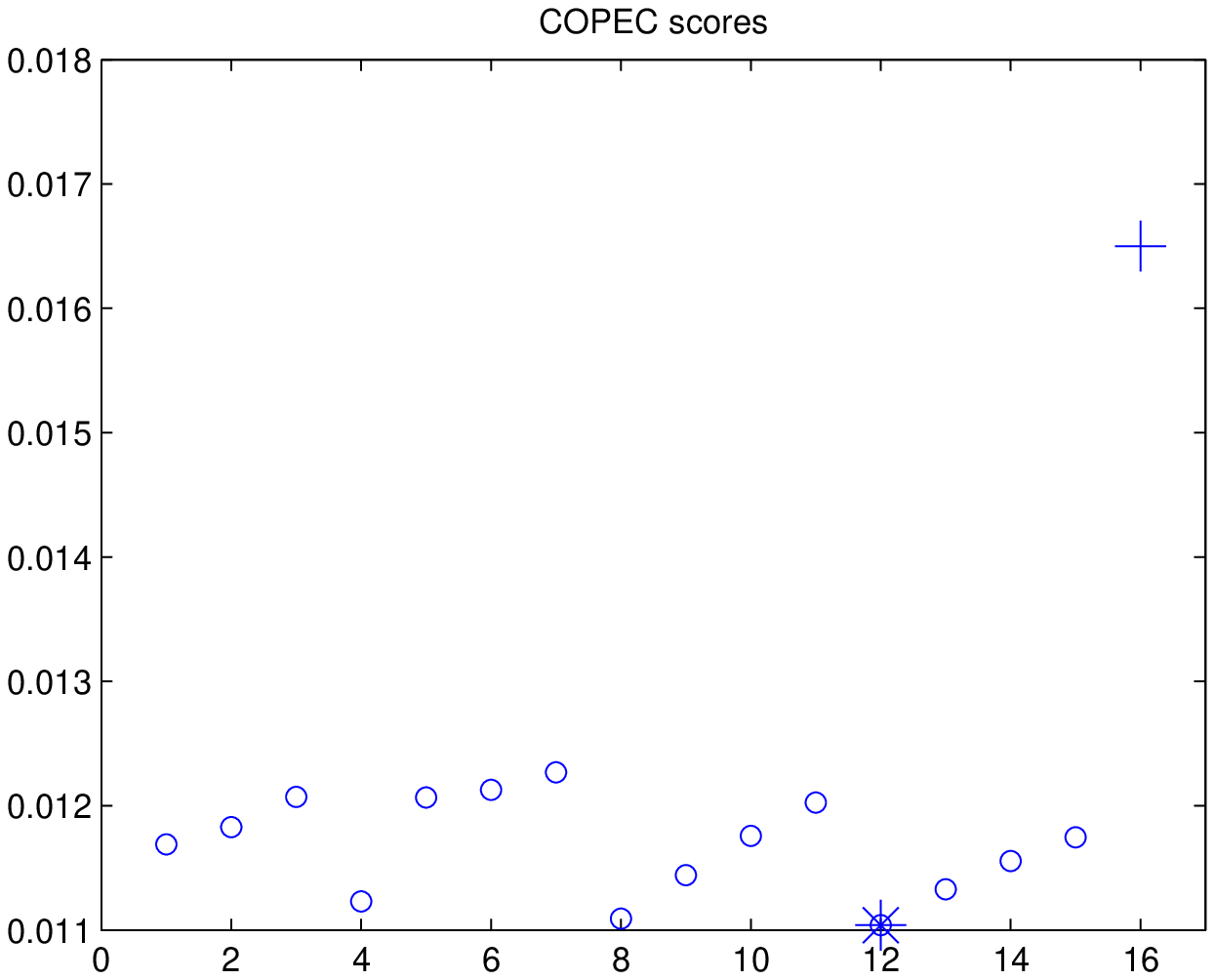}\\
\hspace*{3.5cm}\includegraphics[height=5.8cm]{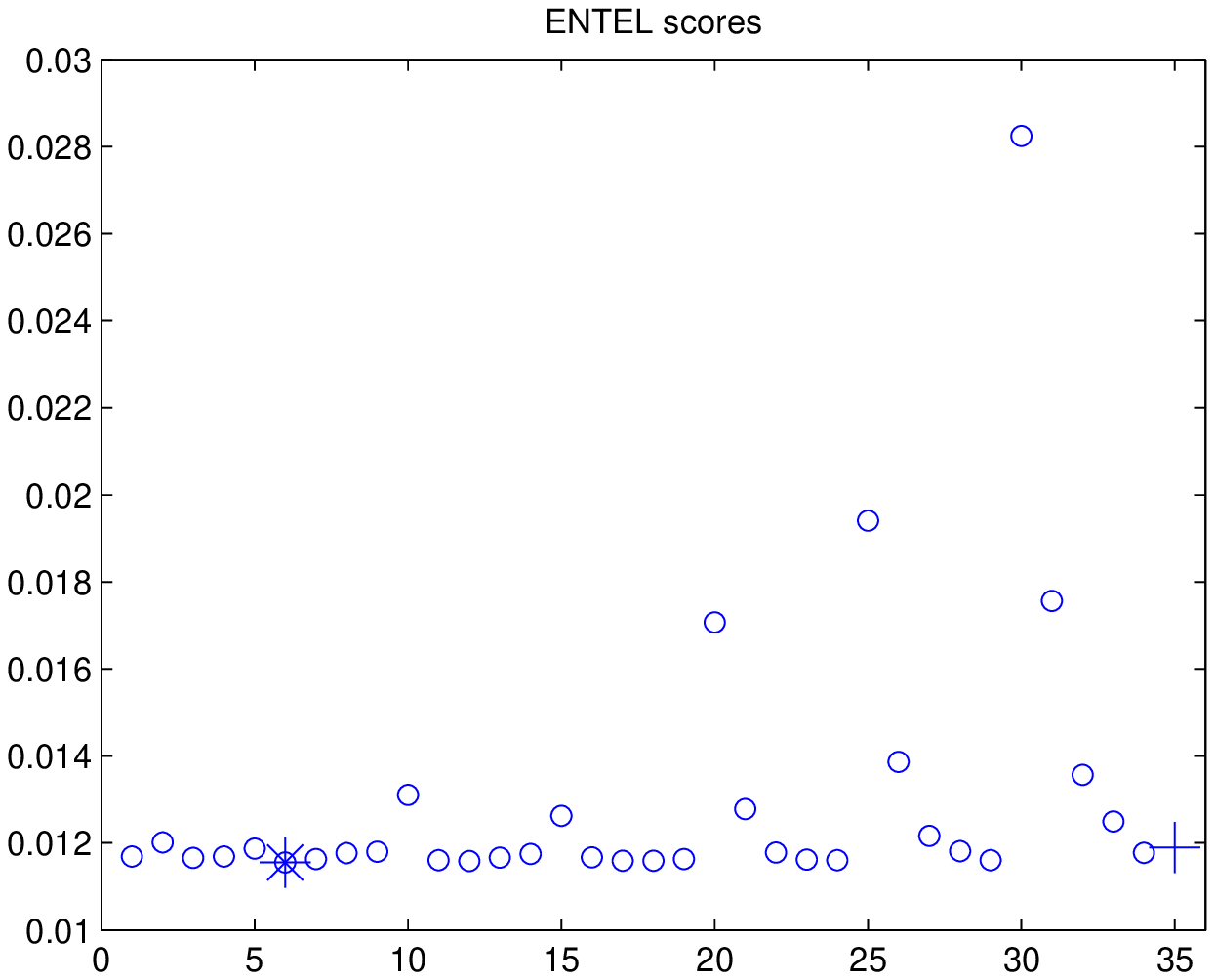}
\label{fig1}
\end{figure}

 Figure \ref{fig1} reports the scores of all
partitions explored by our algorithm and the score of the
partition selected by Quintana et al. (2005a). For the Cementos,
Concha y Toro and Copec shares many partitions, explored by our
algorithm, present a lower value of the score function than the
one selected by Quintana et al. (2005a).

\begin{center}
FIGURE \ref{fig1}  ABOUT HERE \end{center}

\begin{figure}\caption{Concha y Toro: Bayesian linear regression lines. The dashed line represents the regression line obtained considering
only one cluster. Continuous lines are the Bayesian regression
lines for each cluster identified by QIB (first row) and DMT
(second row).
\medskip}
\centering{\includegraphics[height=0.40\textheight]{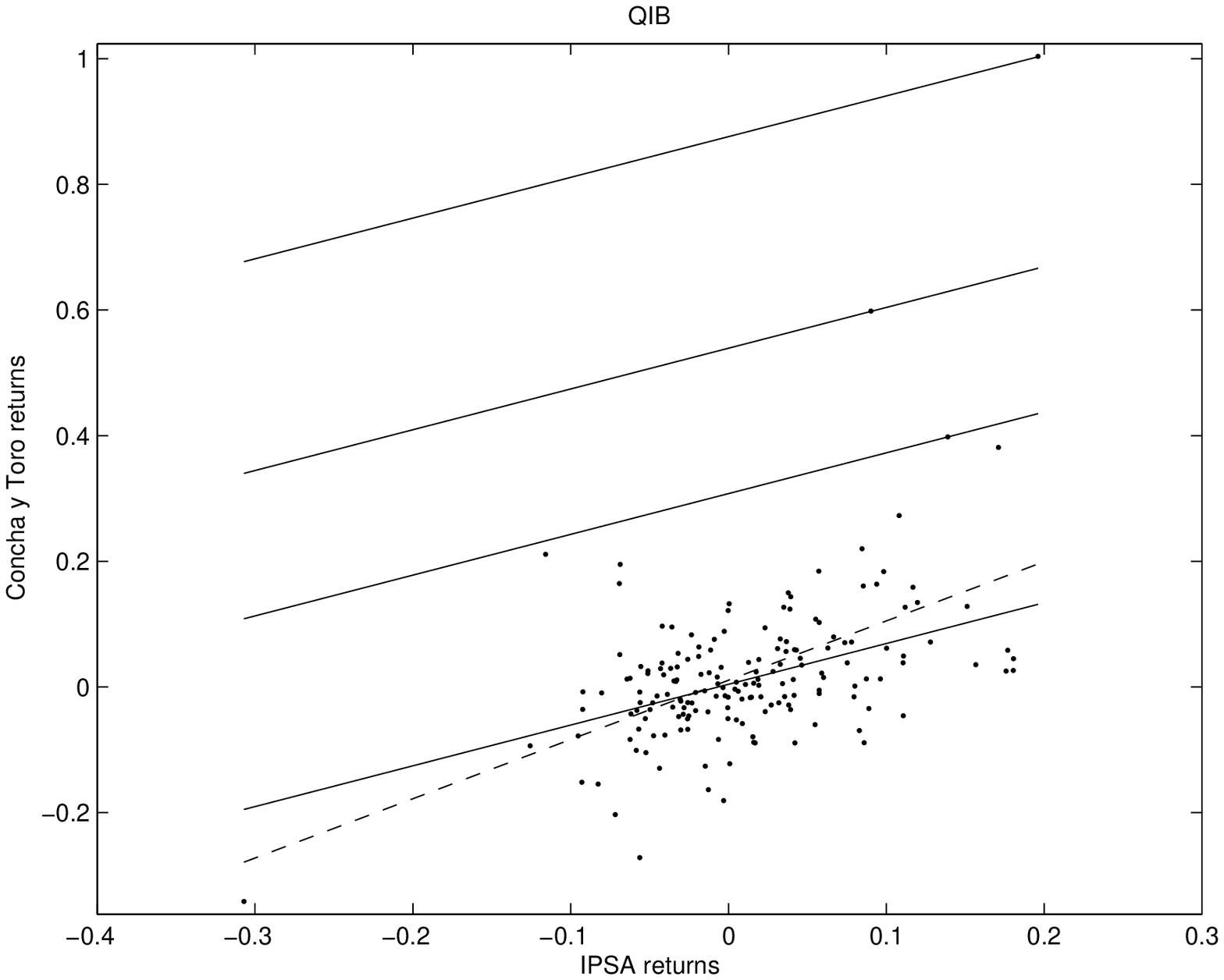}\\
\includegraphics[height=0.40\textheight]{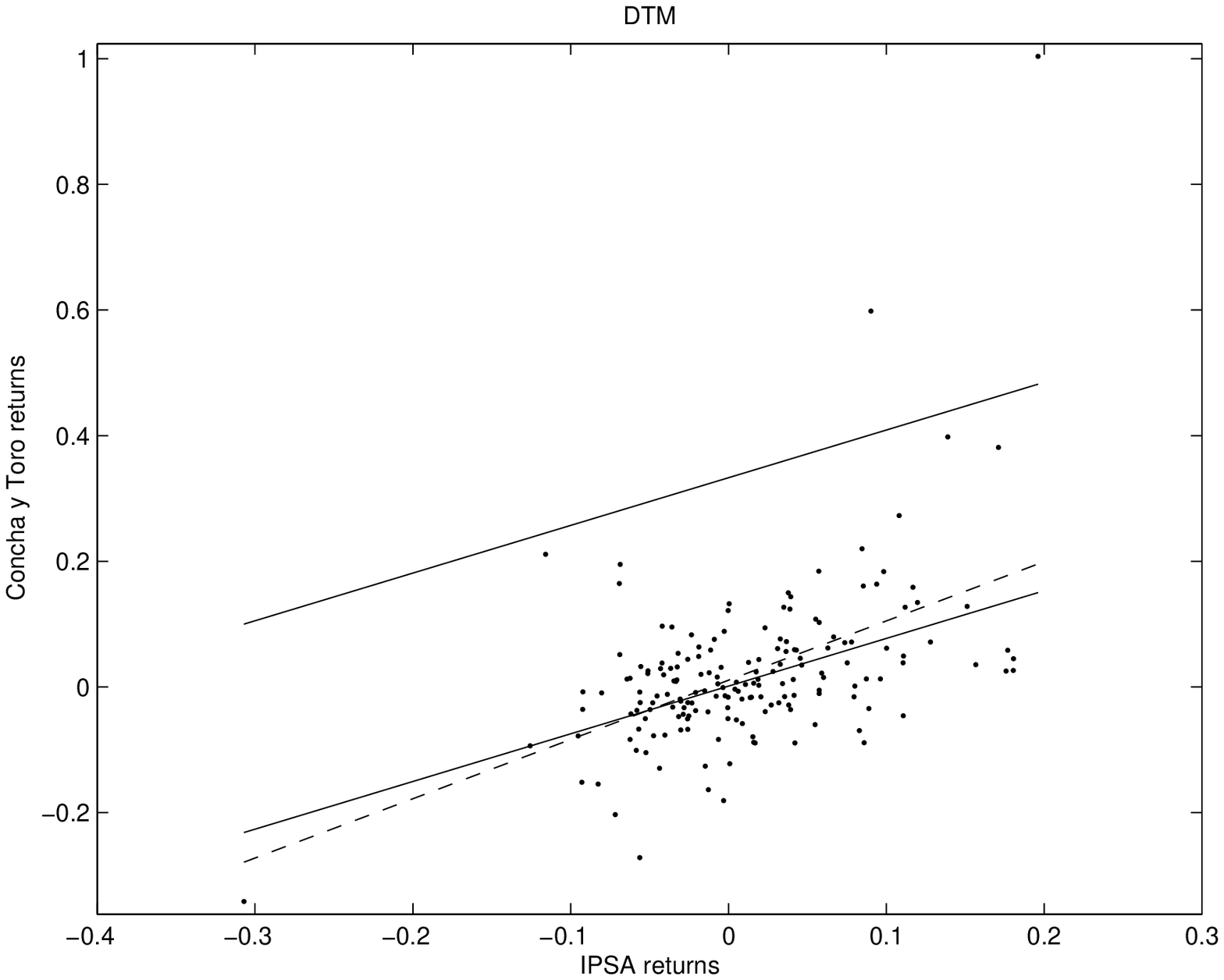}}
\label{fig2}
\end{figure}

 A more detailed analysis of the Concha y Toro share is
provided in Figure \ref{fig2} and Table \ref{Tab2}. In Figure
\ref{fig2} we represent the Bayesian linear regression lines
obtained applying the algorithm by Quintana et al. (2005a) and the
one proposed here. The best partition of Quintana et al. (2005a)
produces a regression line for each detected outlier.
 In Table \ref{Tab2} we
report the Bayesian estimates of the systematic risk under the
three different partition structures considered in Figure
\ref{fig2}.
\begin{center}
FIGURE \ref{fig2}  ABOUT HERE \end{center}

\begin{center}
TABLE \ref{Tab2}  ABOUT HERE \end{center}

In Table \ref{Tab3} we report,  with reference to the Concha y
Toro data, a sensitivity analysis of the results for different
choices of the constant $c$ in (\ref{eqcohesion1}). Note that, for
a wide range of values of $c$ our results are remarkably robust.

\begin{center}
TABLE \ref{Tab3} ABOUT HERE \end{center}

\subsection{Microeconomic analysis}

We performed a microeconomic analysis of the companies under
study, and we list  some events that could have produced the
abnormal behaviour identified by the outliers. All the information
provided   is freely available on the World Wide Web.

\medskip

\noindent 1) CEMENTOS B\'IO-B\'IO S.A.\ \ The Cementos B\'io B\'io
S.A.\ is a company  involved in the production and sale of cement
and lime products, wood and its by-products, premixed concrete and
ceramics.

\noindent In 1992 (outlier 27) it opened a new cement  plant in
Copiapo.

\noindent In  1998 (outlier 107) it expanded the cement plant in
Antofagasta and  started up a  new cement plant in Curic\'o.

\noindent In 1999 (outliers 112, 113) Cementos de Mexico, the
world's third-largest cement manufacturer, entered the Chilean
market by  acquiring  12\% of the Cementos B\'io B\'io S.A.
shares.

\medskip

\noindent 2) CMPC\ \ The  group's principal activity is
manufacturing pulp and paper in Chile. It is an integrated company
that undertakes its industrial work through five business
affiliates (CMPC Celulosa, CMPC Papeles, CMPC Productos de Papel,
CMPC Tissue,  and Forestal Mininco), and owns industrial plants in
Chile, Argentina, Peru and Uruguay.

\noindent The years  from 1990  to 1992 (outlier 15) were
characterised by an expansion in Latin America. In 1990  CMPC
entered  Argentina by purchasing (in partnership with Procter \&
Gamble), Quimica Estrella San Luis S.A.\ (now Prodesa), a
manufacturer of sanitary napkins and paper diapers.  In 1992 CMPC
formed a strategic alliance with Procter \& Gamble to develop
markets for the aforementioned products in Chile, Argentina,
Bolivia, Paraguay, Peru, and Uruguay.

\medskip

\noindent 3) CONCHA Y TORO\ \ Concha y Toro  is  one of the
leading producers of wine in  Chile. It produces and exports a
wide range of wines. In 1994, Concha y Toro became the first
Chilean winery to be listed on the New York Stock Exchange.

\noindent During the years 1991-1993 (outliers 14, 18, 21, 22, 27)
important changes took place.  Concha y Toro  tripled the size of
its vineyards to reduce dependence on outside grape growers and
enrolled the help of French and Californian oenologists. It
modernized its production and transformed the original Concha y
Toro mansion into the head quarters of the firm for its export
operations.

\noindent In 1996  Concha y Toro purchased a vineyard in the
Mendoza region in Argentina. In 1997 the company and the French
firm Baron Philippe de Rothschild S.A.\ endorsed a joint venture
with the aim of producing a wine to the standards of the French
Grand Cru Class�. In  1998 (outlier 97) Concha y Toro launched
Vina Almaviva into the market. In the same year the company ranked
second among wine exporters to the United States.

\medskip

\noindent 4)  COPEC S.A. \ \
  Copec S.A.\ is a diversified Chilean financial holding
company that participates through subsidiaries and related
companies in different business sectors
 (energy,
forestry, fishing, mining and power industries).

\noindent In 1992 (outlier 31) it  united two fisheries to form
Igemar that became the biggest fishing and fish-processing company
in Chile.

\noindent In 1998 (outlier 107) it became Chile's largest exporter
outside of the mining sector.

\noindent In 1999 (outlier 111) COPEC created Air Bp Copec S.A. to
commercialise fuels for national and international air lines, in
joint venture with BP Global Investments.

\medskip

\noindent 5) ENTEL\ \ Entel  was created in 1964 as a state
company, and it was
 privatised in 1986. The group's principal activities are
providing telecommunication services. It also operates in Central
America and Peru aside from its centre of major operations which
is located in Chile.

\noindent In 1996 (oulier 65) Telecom Italia acquired a $19.99\%$
of  Entel shares.

\section{Concluding remarks}

In this paper we  presented  a model for robust inference in
 CAPM  in the presence of outliers. Working in a Bayesian decision
framework, we developed a constrained optimization algorithm for
outlier detection.   Diffe\-ren\-tly from the methodology proposed
by Quintana et al.~(2005a) it appeared to be successful in the
identification of masked outliers and led to  partitions with  a
lower value of the score function.

The outlier identification procedure proposed by Quintana and
Iglesias (2003) is based on a hierarchical divisive method. Their
procedure works by detaching, one by one, the most outlying
component from the
 vector $\hat{\boldsymbol{\alpha}}_{i}^B(y)$ of the Bayesian estimates. This procedure is irreversible,
  that is  once a point is classified in a specific cluster it
 is not
 taken
 any more under consideration.
On the other hand our algorithm    allow  groups of observations
to be considered simultaneously as potential outliers.
  This could be a possible
 explanation of why in some cases, as  for the Concha y Toro share, the algorithm by Quintana and Inglesias (2003)  identifies a
 smaller set of outliers, incurring in the
the masking problem.

A microeconomic analysis is provided to  confirm that the selected
outlying points are linked to extraordinary  events in the history
of the examined companies.

\section*{Acknowledgements}

The authors acknowledge Fernando Quintana for helpful discussion
regarding the computational algorithm and Manuel Galea-Rojas for
providing the IPSA data. We thank Pierpaolo Uberti for comments on
the preliminary version of this paper. We are also grateful to the
associate editor and the referee for valuable comments. The
research of the three authors was (partially) supported by
University of Pavia. The research of the first author was also
(partially) supported both by MUSING (contract number
 027097); the research of the third author was also (partially)
supported by both MIUR, Rome (PRIN 2005132307).

\section*{Appendix: A Gibbs sampling algorithm \label{apA}}

We adapt to our problem a Gibbs sampling algorithm, proposed by
Bush and MacEachern~(1996).
 Consider a generic asset $i$. Given
the starting values $\alpha_0,\ \beta_0$ and $\sigma^2_0$ we
iteratively sample from the following distributions
\begin{eqnarray}
&\;& \beta_i|\sigma_i^2,\boldsymbol{\alpha}_i, \mathbf y_i \sim N
\left\{\frac{b/\gamma_0^2+
\sum_{t=1}^T(y_{i_t}-\alpha_{i_t})x_t}{1/\gamma^2_0+
\sum_{t=1}^T x_t^2},\frac{\sigma_i^2}{1/\gamma^2_0+\sum_{t=1}^T x_t^2} \right\}\; \; \label{fc1}\\
&\;&\sigma_i^2|\boldsymbol{\alpha}_i,\beta_i, \boldsymbol y_i \sim
IG\left\{v_0+\frac{T+|\rho_i|+1}{2},\lambda_0+\frac{(\beta_i-b)^2}{2\gamma_0^2}+\frac{1}{2\tau_0^2}\sum^{|\rho_i|}_{d=1}
(\alpha^*_{i_d}-a)^2 \right. \nonumber\\
&\;&\;\;\;\; \;\;\;\;\;\;\;\;\left.+\frac{1}{2}\sum_{t=1}^T(y_{i_t}-\alpha_{i_t}-\beta x_t)^2\right\}\; \; \label{fc2}\\
&\;&  \alpha_{i_t}| \boldsymbol
\alpha_{i_{-t}},\beta_i,\sigma_i^2, \boldsymbol y \propto \sum_{j
\neq t} \exp \left\{-\frac{1}{2
\sigma^2}(y_{i_t}-\alpha_{i_j}-\beta_i x_t)^2 \right\}
\delta_{\alpha_{i_j}}(\alpha_{i_t}) \nonumber\\
&\;&\;\;\;\; \;\;\;\;\;\;\;\;+\frac{\exp\left\{-(y_{i_t}-\beta
x_t-a)^2/2\sigma_i^2(1+\tau_0^2)\right\}}{\sqrt{1+\tau_0^2}}\;\;
N\hspace{-0.1cm}\left(\frac{y_{i_t}-\beta_i
x_{t}+a/\tau_0^2}{1+1/\tau_0^2},\frac{\sigma_i^2}{1+1/\tau_0^2}\right)
\nonumber
\end{eqnarray}
where $\boldsymbol
\alpha_{i_{-t}}=(\alpha_{i_1},\ldots,\alpha_{i_{t-1}},\alpha_{i_{t+1}},\ldots,\alpha_{i_T})^{\prime}$
and $\delta_{\alpha_j}(\cdot)$ is the Kronecker delta function.

Note that $\beta_i$ and $\sigma_i^2$ are sampled from the
corresponding full conditional whereas each $\alpha_{i_t}$ is
sampled from a mixture of point masses and a normal distribution.
In this way we automatically update both the vector $\boldsymbol
\alpha_i$ and the partition structure.

 Before proceeding to the next Gibbs iteration,
we update the vector   $\boldsymbol \alpha_i$ given the partition
$\rho_i$ sampling  from
\begin{eqnarray}&\;&\alpha^*_{i_d}\sim N
\hspace{-0.1cm}%\left\{
\left(\frac{\sum_{t \in S_d}(y_{i_t}-\beta_i
x_t)+a/\tau_0^2}{|S_d|+1/\tau_0^2},\frac{\sigma_i^2}{|S_d|+1/\tau_0^2}\right)
%\right\}
\; \; \;d=1,\ldots,|\rho_i|. \label{fc3}
\end{eqnarray}

This last step was introduced in Bush and MacEachern~(1996) to
avoid being trapped in sticky patches in the Markov Space.

If the partition structure $\rho_i$ is fixed $\beta_i$,
$\sigma_i^2$ and $\boldsymbol \alpha_i$ are directly sampled from
the corresponding full conditional distributions, (\ref{fc1}), in
(\ref{fc2}) and (\ref{fc3}) respectively.

\begin{sidewaystable}[h!]
\caption{IPSA stock market data:  comparison of the results
obtained via LTS, QIB's  algorithm (Quintana et al.,~2005a) and
DMT's algorithm (De Giuli, Maggi and Tarantola). The second, third
and fourth columns display the clustering structure of the
outliers.
 The last
three columns show the scores of the  partitions selected by
Quintana et al.~(2005a)   and De Giuli, Maggi and Tarantola and
production reduction in score (PRS). {Figures
   in brackets are Monte Carlo
standard errors $\times 10^6.$} }
\begin{small}
\begin{center}
\begin{tabular}{|l|l|l|l|c|c|c|}
  \hline
  Society & $\mathrm{LTS}$ & QIB & DMT & $\mathrm{SC}_\mathrm{QIB}$ & $\mathrm{SC}_\mathrm{DMT}$& PRS\\
  \hline
  CEMENTOS & $\{12, 14, 18, 19, 21, 27, 37, 47, 59, 88,$ & $\{12, 18, 27, 37, 112, 121\},$ & $\{12, 18, 21, 27, 37, 47, 59,$ & 0.0277 (4.79) & 0.0113 (4.59)& 0.4115\\
   B\'{I}O-B\'{I}O S.A.& $91, 107, 112, 113, 121, 133, 161\}$ & \{21, 113\},$\{47, 59, 133\},\{107\}$ & $ 107, 112, 113, 121, 133\}$ & &  &\\
   \hline
 CMPC & $\{3, 15, 49, 50, 54, 57, 100, 112\}$& $\{15\}$ & $\{15\}$ & 0.0122 (45.09)& 0.0122 (45.09) &0.0000\\ \hline
CONCHA Y& $\{12, 14, 18, 21, 22, 23, 27, 29, 32, 83,$ & $\{14\},\{21\},\{27\}$ & $\{14, 18, 21, 22, 27, 97\}$ &  0.0305 (5.44)& 0.0116 (12.57) & 0.6197\\
 TORO & $97, 104, 106, 110, 144\}$ &  &  &   &  &\\
   \hline
  COPEC S.A.& $\{31, 105, 107, 108, 109, 111\}$& $\{31, 107\},\{111\}$ & $\{31, 107, 111\}$ & 0.0165 (13.45)& 0.0110 (12.27) & 0.3333\\
 \hline
 ENTEL & $\{14, 27, 29, 48, 59, 63, 65, 106, 111, 112\}$ & $\{65\}$ & $\{59, 65\}$ & 0.0119 (41.73)& 0.0116 (54.26) & 0.0252\\
\hline
\end{tabular}
\label{Tab1}
\end{center}
\end{small}
\end{sidewaystable}

\begin{table}
\caption{Concha y Toro: Bayesian Estimates of the systematic risk
obtained with only one cluster, with the partition structure of
QI, and with the partition struture of DMT respectively. Figures
   in brackets are Monte Carlo
standard errors $\times 10^5.$}
\begin{center}
\begin{tabular}{|c|c|}
  \hline
  Partition structure & $\hat{\beta}^B$ \\
  \hline
  $S^{0}$ & 0.9430 (9.2638) \\ \hline
QIB   & 0.6486 (7.4916) \\ \hline DMT &  0.7590 (9.0300) \\
  \hline
\end{tabular}\label{Tab2}
\end{center}
\end{table}

\begin{table}
\caption{Concha y Toro: Sensitivity analysis }
\begin{center}\begin{tabular}{|l|l|} \hline
 $c$ in equation (\ref{eqcohesion1})& Outliers \\
\hline
$0.01$, $1$ and $5$ & $\{14, 18, 21, 22, 27, 97\}$ \\
\hline
$10$ and $50$ &  $\{14, 21, 27, 97\}$ \\
\hline
\end{tabular}\label{Tab3}
\end{center}
\end{table}

\newpage

\end{document}